\documentstyle[twoside,fleqn,espcrc2,epsf]{article}

\newcommand{\Dslash}{$D$\kern-0.6em \hbox{/}}
\newcommand{\la}{\raise.16ex\hbox{$\langle$}}
\newcommand{\ra}{\raise.16ex\hbox{$\rangle$}}

\title{Local topological and chiral properties of QCD}

\author{
Ph. de Forcrand\address{SCSC, ETH-Z\"urich, CH-8092 Z\"urich, Switzerland}, 
M. Garc\'ia P\'erez\address{Dept. F\'{\i}sica Te\'orica, Universidad
Aut\'onoma de Madrid, E-28049 Madrid, Spain}, 
J.E. Hetrick\address{Physics Dept., University of the Pacific, 
Stockton, CA 95211, USA},
E. Laermann\address{Facult\"at f\"ur Physik, Universit\"at Bielefeld,
33501 Bielefeld, Germany},
J.F. Lagae\address{HEP Division, Argonne National Laboratory, 
9700 South Cass Avenue, Argonne IL 60439, USA}, 
and 
I.O. Stamatescu\address{FESt, Schmeilweg 5,D-69118, Heidelberg, Germany}}

\begin{document}

\begin{abstract}
To elucidate the role played by instantons in chiral symmetry
breaking, we explore their properties, in full QCD, around the
critical temperature. We study in particular spatial correlations
between low-lying Dirac eigenmodes and instantons. Our measurements
are compared with the predictions of instanton-based models.
\end{abstract}
\maketitle

We have examined the local topological structure of
full QCD and its impact on the physics of chiral symmetry breaking and
restoration, by comparing the local topological structure obtained by
improved cooling\cite{impcool}, with the lowest eigenmodes of the Dirac
operator on the same (uncooled) lattices. We use a set of dynamical
finite temperature $N_t = 12$ (MILC\cite{MILC}) lattices spanning the
chiral phase transition, and have extracted the lowest 8 eigenmodes
and eigenvalues of the staggered Dirac operator.

To summarize our findings, we see:

$\bullet$ agreement with the Banks-Casher relationship using
the density of eigenvalues near zero (not discussed in this writeup).

$\bullet$ good correlation ($\sim$ 70\% level) between the spatial
structure of zero modes and instantons.

$\bullet$ some space-time asymmetry in the local topological
susceptibility above $T_c$ (and not below), the underlying mechanism of
which is under further investigation.

\section*{1. LOCAL STRUCTURE}
%: {\bf $\bar{\psi}\psi(x)$  -vs- $F\tilde{F}(x)$}}
%........................................................................

To illustrate the relationship between instantons and the zero modes
of the Dirac operator we show in Figures 2 and 3, the topological
charge density $F\tilde{F}(x)$, on a configuration obtained
after 150 sweeps of improved cooling. The timeslice shown
happens to contain part of an instanton and anti-instanton, and we
plot isosurfaces of positive and negative values of $F\tilde{F}(x)$.

{\it On this cooled configuration}, we identify the lowest eigenmode of
the Dirac operator; an isosurface of the magnitude of the eigenvector
is plotted along with $F\tilde{F}(x)$, and is shown in Figure 2. We
see that on this smooth configuration in which the UV fluctuations
have been removed, $|\bar{\psi}\psi(x)|$ follows $F\tilde{F}(x)$
exactly, showing that on continuum like configurations, the zero mode
``tracks'' the topology, as expected from continuum arguments.
% (see 
%\cite{SmitVink}, for example).

Next, Figure 3b compares the {\em uncooled} zero mode to the {\em
cooled} topological charge; we see surprisingly good correlation, even
after many (150) cooling sweeps. Cooling identifies the dominant
instanton---anti-instanton (I-A) pairs and the ensemble correlation between
the {\em uncooled} Dirac mode and cooled topological charge density is
about 70\%, after 150 sweeps of improved cooling, on configurations
containing one or more I-A pairs. 
This validates a posteriori the improved cooling process; 
the topology
seen by the Dirac zero mode is largely the same as that seen by the
cooled topological charge. 
% This statement needs further qualification
% though and is under study.

%\vfil\eject

In Figure 3a, we show isosurfaces of $\bar{\psi} \gamma_5 \psi(x)$,
which takes value $\pm 1$ for a right- or left-handed eigenmode respectively.
Yellow and blue surfaces indicate right and left-handedness.
We see what amounts to the chirality flip as quarks ``tunnel''
from an instanton to an anti-instanton. 
%An isosurface is drawn in
%yellow on the right handed component of the zero-mode, and a similar
%isosurface in blue for the left handed component of $\psi_0(x)$

%%FIG:  $q(x)$ and $\bar{\psi}(x) \gamma_5 \psi(x)$
%
%\begin{figure}[htb]
%\centerline{
%\epsfxsize=8cm 
%%\epsfclipoff
%\epsffile[17 61 575 682]{snap0_chi.ps}
% }
%\label{FIG8}
%\caption{Chirality flipping as the quarks tunnel from instantons (red)
%to anti-instantons (blue). The chirality of the eigenmode,
%$\bar{\psi}(x) \gamma_5 \psi(x)$, is either yellow (+) or light blue (-).}
%\end{figure}

%........................................................................
\section*{2. ABOVE {\bf $T_c$}}
%........................................................................

At $T \sim 1.25 ~T_c$, all configurations have Q=0, as $\chi \rightarrow 0$
sharply.  It is very difficult to say that there are instantons and
anti-instantons. After a minimal degree of smoothing ($\sim$20 cooling
sweeps) a fit to non-interacting (anti)instantons or calorons does not
converge. Before that, the configurations are too rough to attempt any
topological identification, and we cannot really say for sure that
there are instantons at all.

We can however try to characterize the fluctuations of
$F\tilde{F}(x)$. In instanton liquid models\cite{Shuryak},
instanton---anti-instanton dipoles are thought to orient in the
timelike direction. Inspired by this theoretical picture we examined
our data in the following way\cite{IlgShur}.

We slice the lattice up into smaller sublattices, taking slabs of a
given thickness $d$. While the slabs have a given thickness in
direction $\mu$, they span the lattice in the other three
directions. $\mu$ is chosen either along the time axis or one of the
spatial axes, so that our slabs span either space or time. Within the
slab we compute $\int [F\tilde{F}(x)]^2 d^4x = Q^2_{\rm slab}$, then
average over all possible slabs of thickness $d$ which are of the same
orientation w.r.t. space or time.

If instantons and anti-instantons are randomly placed in an isotropic
lattice, we expect that, up to a volume normalization factor, spatial
and temporally oriented slabs of the same thickness $d$ should have
the same value of $\langle Q^2 \rangle_d$. This is what we see on the
lattices below $T_c$.

If dipole pairs have formed in the timelike
direction, we would expect $\langle Q^2 \rangle_d$ to plateau for
large $d$ for slabs which span the time-like direction ($\mu$ in the
spatial direction). This is because we are always adding a $Q=0$ pair
to the slab as we increase its thickness, and this does not increase
$Q^2$ in the slab. For spatially spanning slabs on the other hand, the
addition of charge is more or less random.

What we see above $T_c$, after a small amount of cooling (less than 20 sweeps),
is the behavior just described: a plateau in
$\langle Q^2 \rangle_d$ for $d$=spacelike oriented slabs (time-spanning),
and no such plateau in the corresponding $d$=timelike (space-spanning)
slabs. Although it is quickly washed out by cooling, the asymmetry is 
quite clear and unmistakable. What is less
clear is its interpretation.
\vskip 1cm
%FIG:  Asymmetry of $x - t$ slices.
\begin{figure}[htb]
\centerline{
\epsfxsize=6.8cm 
%\epsfclipoff
\epsffile[50 110 582 585]{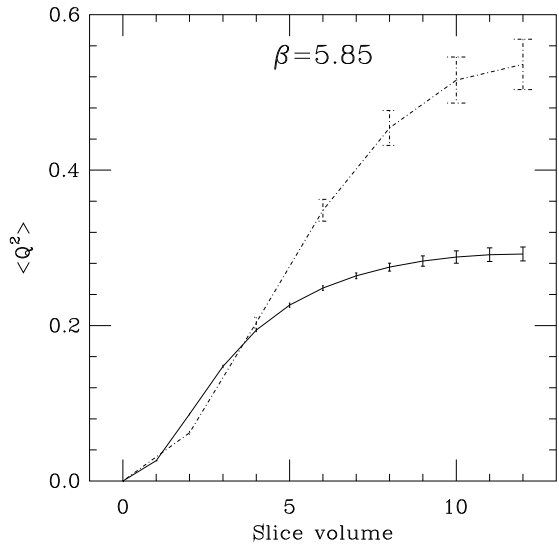}
 }
\label{FIG4}
\vskip -1cm
\caption{$\langle Q^2 \rangle$ versus volume: The lattice is sliced
into sections of given volume, and $\langle Q^2 \rangle$ is computed
on these. The solid curve has time-spanning slices 
($d$ in the spatial direction)
and the dashed curve has space-spanning slices
($d$ is in the timelike direction). If there are dipoles in a
particular direction (timelike it would appear), then $\langle Q^2
\rangle$ should plateau when slices are taken so that they span that
direction.  Notes: the spatial axes have been rescaled to match the
temporal dimension ($N_t = 12$), and $Q^2$ is measured after 5 cooling
sweeps. Below $T_c$, both curves are identical within
statistics.}
\end{figure}

It is important to note that we see this asymmetry for quenched SU(2)
as well, where it is {\it not} predicted to be by instanton liquid models.

%We do not see such an asymmetry in the chiral zero modes. 

Since obtaining this result we have generated synthetic configurations
with a controlled topological charge density, and computed $\langle
Q^2 \rangle_d$ on these. On synthetic configurations we see:

$\bullet$ For randomly placed point charges, $q(x) = \sum_i \pm \delta(x
- x_i)$, there is no asymmetry. This corresponds to the data below $T_c$. 

$\bullet$ With {\em randomly oriented dipoles} having point charges,
we see a trivial asymmetry of geometric origin. 
There is a plateau in both space and time slices,
and the ratio of the plateau values is equal to the ratio of the slab sections
(i.e. 2).

$\bullet$ For randomly placed {\em calorons}, we see a {\bf reverse}
asymmetry which shows dependence on the radius of the caloron. By
reverse asymmetry, we mean that it is the space-spanning ($d$ timelike)
slabs which plateau at a lower value, the reverse of what is seen in
Figure 1.
\vskip -1.9cm

\begin{figure}[htb]
\centerline{
\epsfxsize=7cm 
%\epsfclipoff
\epsffile[32 135 575 772]{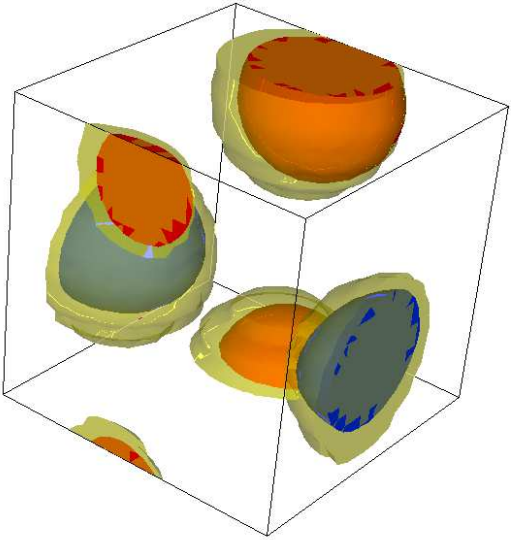}
 }
\label{FIG1}
\vskip -0.5cm
\caption{Isosurfaces of topological charge density $F\tilde{F}(x)$
(red[+] and blue[-]) and $\bar{\psi_0}(x) \psi_0(x)$ (yellow) using the
lowest eigenmode $\psi_0$. Here $\bar{\psi_0}(x) \psi_0(x)$ is
computed on the cooled configuration.}
\vskip -0.5cm
\end{figure}

%\begin{figure}[htb]
%\vskip -3.8cm
%\centerline{
%\epsfxsize=7.5cm 
%%\epsfclipoff
%\epsffile[0 135 575 772]{snap0_chi_ff.ps}
% }
%\vskip -1cm
%\centerline{
%\epsfxsize=7.5cm 
%\epsfclipoff
%\epsffile[0 135 575 772]{snap0_ff.ps}
% }
%\label{FIG2}
%\vskip -0.5cm
%\caption{Isosurfaces of topological charge density $F\tilde{F}(x)$
%(red[+] and blue[-]) and $\bar{\psi_0}(x) \psi_0(x)$ (yellow) using the
%lowest eigenmode $\psi_0$. Here $\bar{\psi_0}(x) \psi_0(x)$ is
%computed on the original {\em uncooled} configuration. Above (a),
%chiral components; Below (b) $|\bar{\psi_0}\psi_0(x)|$}
%\vskip -1cm
%\end{figure}

\begin{figure}[htb]
\vskip -2.3cm
\centerline{
\epsfxsize=7.5cm 
%\epsfclipoff
\epsffile[0 135 575 772]{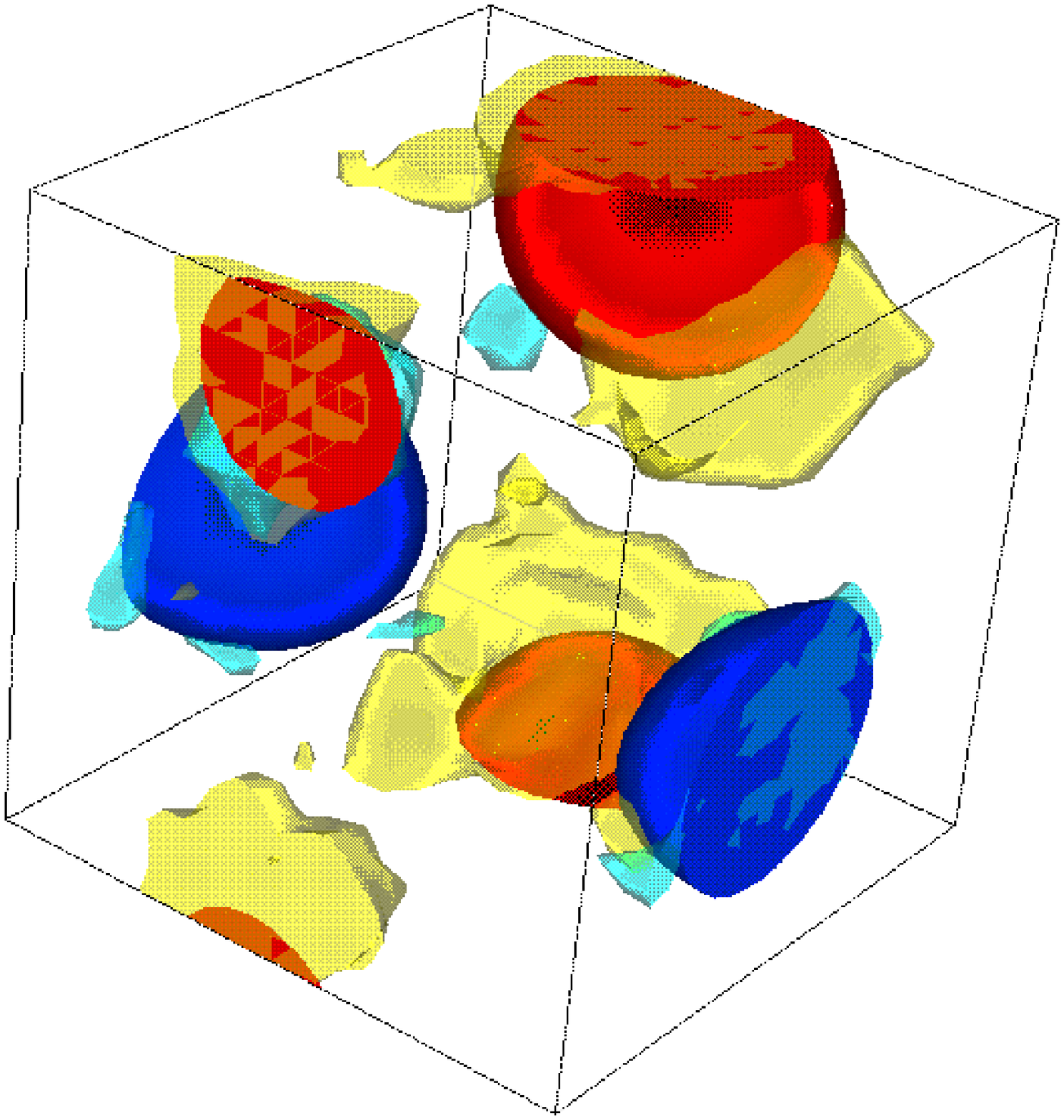}
 }
\vskip -1.3cm
\centerline{
\epsfxsize=7.5cm 
\epsfclipoff
\epsffile[0 135 575 772]{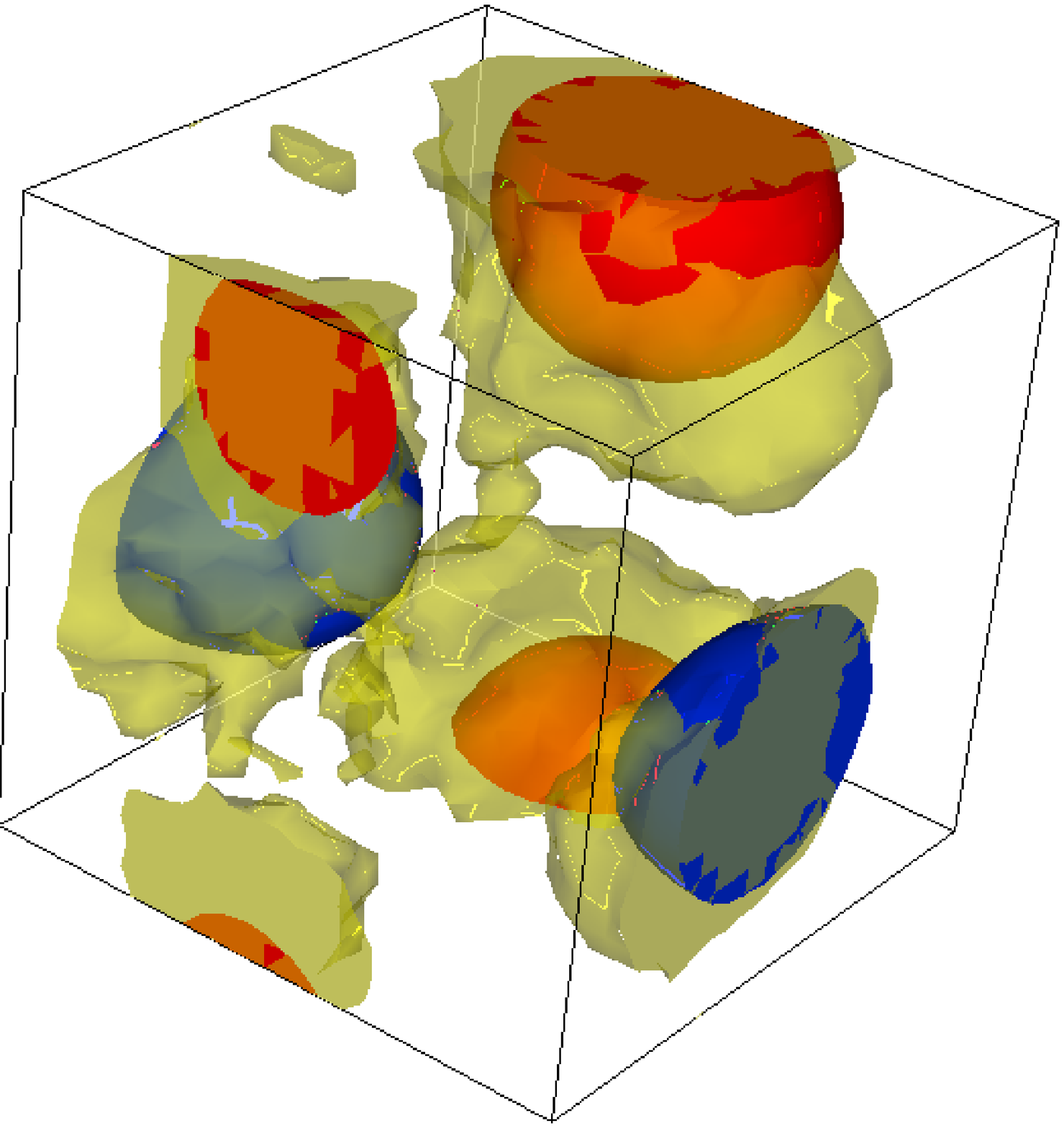}
 }
\label{FIG2}
\vskip -0.6cm
\caption{Isosurfaces of topological charge density $F\tilde{F}(x)$
(red[+] and blue[-]) and $\bar{\psi_0}(x) \psi_0(x)$ (yellow) using the
lowest eigenmode $\psi_0$. Here $\bar{\psi_0}(x) \psi_0(x)$ is
computed on the original {\em uncooled} configuration. Above (a),
chiral components; Below (b) $|\bar{\psi_0}\psi_0(x)|$}
\vskip -1cm
\end{figure}

\end{document}